\documentclass[reprint,floats,floatfix,amsmath,amssymb,nofootinbib,longbibliography]{revtex4-2}
\usepackage{amsfonts}
\usepackage{graphicx}
\usepackage{dcolumn}
\usepackage{bm}
\usepackage{natbib}
\usepackage{wasysym}
\usepackage{multirow}
\usepackage{aas_macros}
\usepackage[hyperindex,colorlinks]{hyperref}

\newcommand{\rs}{\rho^{*}}
\newcommand{\bs}{\boldsymbol}
\newcommand{\bx}{\bar{x}}

\newcommand{\s}{\scriptscriptstyle}

\begin{document}

\title{\textbf{In-depth analysis of violations of the (strong) equivalence principle in scalarized Einstein-Gauss-Bonnet theories} 
}% 

\author{Mart\'in G. Richarte}
\email{martin@df.uba.ar}
\affiliation{Departamento de F\'isica - Universidade Federal do Esp\'irito Santo, 29075-910 Vit\'oria, ES, Brazil}
\affiliation{PPGCosmo, CCE - Universidade Federal do Esp\'irito Santo, 29075-910 Vit\'oria, ES, Brazil}
\affiliation{Departamento de F\'isica, Facultad de Ciencias Exactas y Naturales,
Universidad de Buenos Aires, Ciudad Universitaria 1428, Pabell\'on I, Buenos Aires, Argentina}

\author{Júnior D. Toniato}
 \email{junior.toniato@ufes.br}
\affiliation{Departamento de Química e Física - Centro de Ciências Exatas, Naturais e da Saúde, Universidade Federal do Espírito Santo - Campus Alegre, ES, 29500-000, Brazil.}%
\affiliation{Núcleo de Astrofísica e Cosmologia - Cosmo-Ufes, Universidade Federal do Espírito Santo, Vitória, ES,  29075-910, Brazil.}

\date{\today}% It is always \today, today,

\begin{abstract}
We conducted a theoretical analysis of the violation of the equivalence principle within a broad class of scalar-Gauss-Bonnet theories that exhibit spontaneous scalarization. Beginning with the Jordan frame, we performed a conformal mapping to identify the equivalent model in the Einstein frame. This approach revealed that the Gauss-Bonnet coupling introduces a mixed term that links the Einstein tensor to the kinetic terms, along with an additional kinetic term associated with the box operator, echoing characteristics of Horndeski-like theories. We investigate a potential violation of the weak equivalence principle using a point-like particle framework for dark matter and baryonic matter. Our findings reveal that differing couplings lead these particles to follow distinct geodesics in the Einstein frame. We also examined the violation of the strong equivalence principle through the Nordtvedt effect within the framework of an extended parametrized post-Newtonian (PPN) formalism. We provide a concrete example of scalarized theories featuring extended PPN parameters that deviate from general relativity, comparing these results against observational constraints from the Cassini mission, the MESSENGER mission, and the Lunar Laser Ranging experiment. While the Cassini bounds remain the most stringent, the constraints on the Nordtvedt parameter offer significantly better restrictions on the parameter space than those derived from the precession rate of Mercury’s perihelion.\\

\end{abstract}

\maketitle

\section{Introduction}
 In scalarization mechanisms, a coupling function links a dynamic scalar field to the metric field, which is incorporated into the matter action \cite{Damour:1993hw}. This integration allows their model to closely resemble General Relativity (GR) under weak gravitational conditions and low velocities. However, significant deviations emerge in strong gravitational fields. For example, scalarized models can provide valuable insights into neutron stars when analyzed through the post-Keplerian framework, potentially revealing effects associated with the influence of the scalar field. Over the past decade, a variety of scalarization theories have been extensively studied (see \cite{Doneva:2022ewd} for a comprehensive review).  Moreover,  the exploration of quadratic Gauss-Bonnet theories has gained prominence, particularly in the context of hairy black holes \cite{Antoniou:2017acq, Herdeiro:2018wub, Dima:2020yac, East:2021bqk, Berti:2020kgk, Zhang:2021nnn, Silva:2020omi, Herdeiro:2020wei, Cunha:2019dwb, Lee:2018zym, Lee:2021uis, Antoniou:2024gdf, Antoniou:2024hlf, Belkhadria:2025lev} and compact stellar objects \cite{Silva:2017uqg, Mendes:2018qwo, Kuan:2021lol}.\\

The Einstein-Gauss-Bonnet model coupled to a scalar field (sEGB) has garnered significant attention as an intriguing framework that incorporates two key components \cite{Clifton:2011jh, Koyama:2015vza, Joyce:2014kja, Ishak:2018his, Nojiri:2017, Nojiri:2011}: a curvature correction to the Einstein equations and the introduction of an additional degree of freedom, represented by a scalar field. Various versions of this proposal have been extensively explored in the literature, highlighting its versatility and potential implications \cite{Clifton:2011jh,Fernandes:2020nbq}. The novel features of this model within the context of the PPN formalism have been examined \cite {Clifton:2020xhc,Toniato:2024gtx,Richarte:2025dag}, shedding light on how these modifications may affect different gravitational phenomena. In addition, the scalar EGB model was examined in the context of the black hole-neutron star binary events known as GW200105 and GW200115, leading to the constraint $\sqrt{\alpha_{GB}}< 1.33 \,\rm{km}$ at the $2\sigma$ level \cite{Lyu:2022gdr}. %\\ 

One of the foundational pillars of GR is the various formulations of the equivalence principle. In particular, the strong equivalence principle (SEP) asserts that the results of any fundamental physics experiment remain locally unchanged in the presence of gravitational fields \cite{Will_2018}. Interestingly, Nordtvedt proposed the intriguing idea that gravitational energy could exert distinct influences on inertial and gravitational mass \cite{Nordtvedt:1968qs, Nordtvedt:1968qr}. The following ratio can parametrize such possible deviations,
\[\frac{M_{g}}{M_{\iota}}=1+\frac{E_g}{M_{\iota}c^{2}}\eta_{_{N}},  \]
where $E_g$ represents gravitational energy, $M_g$ and $M_\iota$ are the gravitational and inertial masses, respectively, and $\eta_{N}$ is the Nordtvedt parameter that governs this violation. For instance, using the lunar laser ranging (LLR) method, it has been estimated that $\eta_{N}=(-0.2 \pm 1.1)\times 10^{-4}$ \cite{Hofmann:2018myc}. As demonstrated, binary pulsar systems in the strong-field limit serve as more effective candidates for testing the SEP \cite{Damour:1991rq}, yielding $|\frac{M_{g}}{M_{i}}-1|< \mathcal{O}(10^{-2})$ at the $2\sigma$ level.  Naturally, one might wonder about the constraints on the Nordtvedt parameter within the context of the sEGB models under the PPN formalism. 

Our study centers on the examination of how the strong and weak equivalence principles are violated in the context of scalarized theories. Our objective is to demonstrate how constraints derived from PPN parameters can be used to assess the extent of these violations, which may vary based on the specific dataset and analytical techniques employed. The paper is structured into four main sections. We first discuss the field equations governing the scalarized Einstein-Gauss-Bonnet theories in both the Jordan and Einstein frames. We will also discuss the implications of conformal transformations and the relevant geodesic equations, analyzing how these concepts interplay with the equivalence principles.

In particular, we demonstrate through various approaches that while matter follows geodesics in the Jordan frame, this behavior does not persist upon transitioning to the Einstein frame. Utilizing a point-like particle approach for both dark matter and baryonic matter in conjunction with Eardley's methodology, we show that these particles follow distinct geodesics in the Einstein frame if their couplings are neither universal nor identical. This finding is further illustrated by examining the geodesic equations in the non-relativistic limit and analyzing their relative accelerations. Subsequently,  we will outline the extended PPN formalism and focus on the Nordtvedt frame effect \cite{Nordtvedt:1968qs, Nordtvedt:1968qr} for massive self-gravitating bodies, and its connection with the violation of the SEP.  Later, we review findings that illustrate violations of the SEP in different experiments/missions.

We will examine the degree of violation of the equivalence principles based on the analysis of the extended PPN parameters for sEGB models by performing a joint analysis which combines the constraints from the Casssini and MESSENGER missions, along with the LLR experiment. Finally, we will summarize our key results, emphasizing the implications of our findings for future research in scalarized theories. We adopt geometrized units, where $c=1$ and $8\pi G=1$, unless otherwise noted. Consistently, we will employ the metric signature $(-,+,+,+)$ throughout our analysis.

\section{Scalarized Einstein-Gauss-Bonnet theories}\label{sec:scalarEGB}
\subsection{Jordan frame}
 We explore a generalized scalar-tensor model  within a conformal (Jordan) frame, as discussed in several works  \cite{Damour:1993hw,Faraoni:1999hp, Sotiriou:2008rp}. The action  has two different contributions, the scalarized model  along with the additional matter fields:
\begin{equation}\label{ac1}
    S=S^{J}_{\rm{scal}}[\Phi, g_{\mu\nu}] + S^{J}_{\rm{matt}}[g_{\mu\nu}, \chi].~~
\end{equation} %
Here, the upper index $J$ indicates the Jordan frame, the scalar field is denoted as $\Phi$, $g_{\mu\nu}$ is the physical metric in the Jordan frame, and $\chi$ accounts for additional matter fields.  The Lagrangian for the scalarized models is given by 
\begin{equation}\label{acb}
   {\cal{L}}^{J}_{\rm{scal}}=\frac{1}{2\kappa}M(\Phi)R-Q(\Phi)\nabla_{\mu}\Phi\nabla^{\mu}\Phi +\frac{\varepsilon}{8}f(\Phi){\cal G},
\end{equation}
where  $\kappa = 8\pi G/c^{4}$, $R$ represents the Ricci scalar, ${\mathcal{G}} = R^2 + R_{abcd} R^{abcd} - 4 R_{ab} R^{ab}$ is the topological Gauss-Bonnet term, and the functions $M(\Phi)$, $Q(\Phi)$, and $f(\Phi)$ are arbitrary (coupling) functions of the scalar field, $\Phi$.  The Lagrangian density is constructed by incorporating the four-dimensional volume element, $d\mu_{g}=\sqrt{-g} d^4x $. The parameter $\varepsilon$  serves as a bookkeeping symbol.  
By varying the action with respect to the metric, the following field equation is obtained:
\begin{align}\label{fe}
&M(\Phi) G_{\alpha \beta }- Q(\Phi) \left({\nabla}_{\alpha }\Phi {\nabla}_{\beta }\Phi + X g_{\alpha \beta } \right) \ + \nonumber \\ \nonumber \\
&\varepsilon f'(\Phi) \left[-\frac{R}{2}{\nabla}_{\beta }{\nabla}_{\alpha }\Phi + R_{(\alpha \gamma } {\nabla}^{\gamma}{\nabla}_{\beta)}\Phi - G_{\alpha \beta } \Box\Phi \right]- \nonumber\\ \nonumber \\
&\varepsilon f'(\Phi) \left[ g_{\alpha \beta } R_{\gamma \delta } {\nabla}^{\gamma }{\nabla}^{\delta}\Phi + {R}_{\alpha \gamma \beta \delta }{\nabla}^{\gamma }{\nabla}^{\delta}\Phi \right] \ + \nonumber\\ \nonumber \\
&\varepsilon f''(\Phi) \bigg[-\frac{R}{2}{\nabla}_{\alpha }\Phi{\nabla}_{\beta }\Phi + R_{(\alpha \gamma } {\nabla}^{\gamma}\Phi{\nabla}_{\beta)}\Phi\bigg]+ \nonumber\\ \nonumber \\
&\varepsilon f''(\Phi) \bigg[
2XG_{\alpha \beta } - g_{\alpha\beta}R_{\gamma \delta }{\nabla}^{\gamma}\Phi{\nabla}^{\delta}\Phi +{R}_{\alpha \gamma \beta \delta}{\nabla}^{\gamma }\Phi{\nabla}^{\delta}\Phi \bigg] \ - \nonumber \\ \nonumber \\
&M'(\Phi) \left({\nabla}_{\beta }{\nabla}_{\alpha }\Phi -  g_{\alpha \beta } \Box\Phi \right) - M''(\Phi) {\nabla}_{\alpha }\Phi {\nabla}_{\beta }\Phi + \nonumber \\
\nonumber \\
&\qquad\qquad\qquad\qquad\qquad\qquad 2M''(\Phi)X g_{\alpha \beta } =\kappa T_{\alpha \beta }.
\end{align}
In the above equation, $G_{\alpha\beta}$ represents the Einstein tensor, and
\begin{equation}
    X = -\tfrac{1}{2}\nabla^{c}\Phi\nabla_{c}\Phi.
\end{equation}
In this context, the prime symbol indicates a derivative with respect to the scalar field, while the indices within parentheses signify symmetrization. By taking the trace of Equation (\ref{fe}), we can rewrite the equation as follows:
\begin{align}
  &M(\Phi) R + 2Q(\Phi)X  - \varepsilon f'(\Phi)\Big(\frac{R}{2} \Box\Phi - R_{ab} \nabla^{b}\nabla^{a}\Phi\Big) +\nonumber \\
  &\varepsilon f''(\Phi)\Big(RX + R_{ab} \nabla^{a}\Phi\nabla^{b}\Phi\Big) 
  -3 M'(\Phi)\Box\Phi + \nonumber \\ 
  &6 M''(\Phi) X = - \kappa T,
\end{align}
where $T$ denotes the trace of the energy-momentum tensor (EMT). The generalized scalar field equation is given by 
\begin{align}
 Q(\Phi)\Box\Phi - Q'(\Phi)X +  M'(\Phi)\frac{R}{2} + \frac{\varepsilon}{8} f'(\Phi){\cal G}=0.
\end{align}

Setting $M=1$ and $Q=1$  allows us to recover the standard modified scalar field equation within the scalarized EGB model, as discussed in \cite{Doneva:2022ewd},\footnote{The minimal conditions for spontaneous scalarization in black hole scenarios require that the coupling satisfies  $f'(\Phi_0)=0$ and $f''(\Phi_0){\cal G}<0$ for some $\Psi_{0} \in \mathbb{R}$  \cite{Doneva:2022ewd}.}
\begin{align}\label{esta}
 \Box\Phi+ \frac{\varepsilon}{8} f'(\Phi){\cal G}=0.
\end{align}
In some cases, it can be advantageous to operate in the decoupling limit, where backreaction terms become negligible. This allows the Einstein field equations to be sourced solely by a free scalar field, effectively treating it as the source of the EMT. In this framework, the scalar field evolves according to the equation (\ref{esta}), while all terms that mix the curvature tensor with the kinetic scalar field terms are neglected. This particular scenario proves especially valuable for investigating the descalarization process \cite{Silva:2020omi}.

Even when $M=1$, the equation of motion for the scalar field yields a distinct subset compared to the previous case (\ref{esta}). This difference arises due to the presence of a new type of coupling between the kinetic terms and the scalar field $\Phi$. Such a coupling can alter the dynamics of the scalar field, leading to new physical implications and behaviors that are not captured in the earlier formulation.   \\

\subsection{Einstein frame}
At this point, we will investigate the type of theory that emerges when we perform a conformal transformation to the Einstein frame \cite{Sotiriou:2008rp,Postma:2014vaa}. This transformation typically involves rescaling the metric and adjusting the scalar field accordingly, which can reveal the underlying characteristics of the theory in a more familiar form. 

Let us consider a map $\Gamma: (\mathbf{M}, g^{J})\rightarrow  (\mathbf{M}, \hat{g})$ such that  the metric is re-scaled by a continuous  non-zero real function  $\Omega^{2}(x)$:
\begin{eqnarray}
 g^{J}_{\mu\nu}&=&\Omega^{-2} \hat{g}_{\mu\nu}  \rightarrow    g^{\mu\nu}_{J}=\Omega^{2} \hat{g}^{\mu\nu},\\
 \sqrt{-g^{J}}d^{4}x&=&\Omega^{-4}\sqrt{-\hat{g}}\,d^{4}x.
\end{eqnarray}
The superscript $J$ once again stands for quantities in the Jordan frame, whilst the symbol $~\hat{}~$ refers to quantities in the Einstein frame. It is evident that this mapping is not a coordinate transformation. Moreover, the null-like geodesics remain invariant provided the conformal transformation preserves angles. The Ricci scalar transforms as follows,
\begin{equation}\label{einsJ}
 R^{J}=\Omega^{2}\big(\hat{R} + 3\hat{\Box}\ln \Omega^2 -\frac{3}{2}\hat{\nabla}^{\mu}(\ln \Omega^2) \hat{\nabla}_{\mu}(\ln \Omega^2) \big).
\end{equation}
In Eq. (\ref{einsJ}), we introduce a concise shorthand notation denoting operators with a hat to indicate that the contraction operation incorporates the metric from the Einstein frame.

We begin by analyzing the term with the Ricci scalar field coupled to the $M$-function. We need to identify the essential conditions required to reproduce the standard Einstein-Hilbert term.. The former term can be written as follows,
\begin{eqnarray}
 \frac{\hat{M}(\Phi)}{2\kappa}\sqrt{-\hat{g}}\Big[\hat{R} + 3\hat{\Box}\ln \Omega^2 -\frac{3}{2}\hat{\nabla}^{\mu}(\ln \Omega^2) \hat{\nabla}_{\mu}(\ln \Omega^2) \Big], ~~~~~
\end{eqnarray}
To recover the standard factor multiplying $\hat{R}$, up to a general constant, we have $\hat{M}(\Phi)=1$, or equivalently,  $M(\Phi)= \Omega^{2}$. 
Once the condition above is imposed, we can disregard the term $\hat{\Box}\ln \Omega^2$, since it represents a total derivative. Then, we are left with the standard Ricci term plus a kinetic term in $\ln \Omega^2$.  Besides, there is another kinetic term associated with the $Q$ coupling function, which, after incorporating the contribution from the volume element, can be expressed as
\begin{eqnarray}
-\sqrt{-\hat{g}} \hat{Q}(\Phi) \hat{\nabla}_{\nu}\Phi \hat{\nabla}_{\mu}\Phi,
\end{eqnarray}
where $\hat{Q}(\Phi)= Q(\Phi)\Omega^{-2}$. Under the conformal maps,   
the kinetic term reduces to $ g^{\mu\nu}_{J}\nabla_{\mu}\Phi\nabla_{\nu}\Phi=\Omega^{2} \hat{g}^{\mu\nu}\hat{\nabla}_{\mu}\Phi\hat{\nabla}_{\nu}\Phi$. If we wish to put together both kinetic terms, we must relax another condition by imposing a parametrization of the form $\Phi(\psi)$ and $\Omega(\psi)$, where $\psi$ is a new scalar field. This parametrization allows us to express the dependencies of the kinetic terms on the new scalar field  $\psi$, enabling a unified treatment of the kinetic contributions in the action, yielding $-\frac{1}{2}  \hat{\nabla}_{\nu}\psi \hat{\nabla}^{\mu}\psi $. The relation among these variables is  given by 
\begin{eqnarray}\label{relax}
\frac{d \psi}{d\Phi}=\Bigg[\frac{2Q}{\Omega^2}+\frac{3}{2\kappa}\bigg(\frac{ d\ln \Omega^2}{d \Phi}  \bigg)^2        \Bigg]^{1/2}.
\end{eqnarray}
For a given choice of $Q(\Phi)$ and $\Omega(\Phi)$, it is indeed possible to reconstruct the functional form  $\psi=\psi(\Phi)$ by directly integrating the equation above. 
In summary, the Ricci-scalar term, along with the kinetic factor of the new scalar field, they form the action term,
\begin{eqnarray}
S_{\textbf{R+K}}&=&\int{\sqrt{-\hat{g}} d^{4}x}\Bigg[\frac{1}{2\kappa}\hat{R} -\frac{1}{2} \hat{g}^{\mu\nu} \hat{\nabla}_{\mu} \psi \hat{\nabla}_{\nu}  \psi \Bigg].~~~~~~\label{e+k}
\end{eqnarray}

Before proceeding with the Gauss-Bonnet contribution, it is important to highlight that the matter action takes on a new form following the application of a conformal transformation:
\begin{eqnarray}
\hat{S}_{\rm{matt}}[\Omega^{2}(\Phi)\hat{g}_{\mu\nu}, \chi].
\end{eqnarray}
The inclusion of the latter coupling will lead to several significant changes in the Einstein frame, notably including the non-conservation of the EMT and the SEP. Subsequently, we will discuss the conditions under which the weak equivalence principle (WEP)  may be violated in the Einstein frame, considering more general assumptions.

The transformation rule for the Gauss-Bonnet topological invariant exhibits a complexity that surpasses that of the standard Ricci scalar. This intricacy manifests in the introduction of additional terms that blend curvature contributions with kinetic terms. Moreover, these terms encompass non-linear kinetic features reminiscent of Horndesky-like theories, which are characterized by their rich structure and the interplay between scalar fields and curvature \cite{Toniato:2024gtx,Richarte:2025dag}. Indeed, under a conformal transformation, the Gauss-Bonnet term's behavior is altered in such a way that it no longer retains its straightforward form \cite{Carneiro:2004rt}, namely
\begin{align}\nonumber
\sqrt{-g^{J}} &\gamma(\Phi)\mathcal{G}_{J} = \sqrt{-\hat{g}} \gamma(\Phi)\Big[ 
\hat{\mathcal{G}} 
- 8 \Omega^{-1}\hat{G}^{\mu\nu} \hat{\nabla}_{\mu} \hat{\nabla}_{\nu} \Omega  \\ \nonumber
& -4 \hat{R}\Omega^{-2} \hat{\nabla}_{\mu}\Omega \hat{\nabla}^{\mu}\Omega +8 (\hat{\nabla}_{\nu}\Omega\hat{\nabla}^{\nu} \Omega  \hat{\nabla}_{\mu}\Omega\hat{\nabla}^{\mu} \Omega) \\ \nonumber
&-8(\hat{\nabla}_{\mu} \hat{\nabla}_{\nu}\Omega \hat{\nabla}^{\mu} \hat{\nabla}^{\nu}\Omega)  -24 \Omega^{-3} \hat{\nabla}_{\nu}\Omega\hat{\nabla}^{\nu} \Omega \hat{\Box} \Omega \\
&+24 \Omega^{-4} \hat{\nabla}_{\mu}\Omega\hat{\nabla}_{\nu} \Omega  \hat{\nabla}^{\mu}\Omega\hat{\nabla}^{\nu} \Omega \Big],  
\end{align}
where $\gamma(\Phi)$ is a redefinition of $f(\Phi)$ in terms of eq. (\ref{acb}).\footnote{We can express the coupling as $\gamma(\Phi)=(\varepsilon/8) f(\Phi)$. Here, $\varepsilon$ serves as a booking parameter. However, it is important to note that this expression is quite generic and will be tailored later for our convenience. } 
Until now, the specific connection between the conformal factor and the scalar field has been somewhat open-ended. To establish a clear physical scheme that aligns with the Einstein frame, we need to define this relationship or make some assumptions about it.  In principle, the latter task is not easy to tackle provided  our new scalarization model is much richer than the previous one, as it includes higher-order kinetic terms such as $\hat{\Box} \Omega(\hat{\nabla}\Omega)^{2}$ and $\hat{\nabla}_{\mu}\hat{\nabla}_{\nu} \Omega \hat{\nabla}^{\mu}\hat{\nabla}^{\nu} \Omega$. Additionally, mixed curvature terms involving the Einstein tensor and  the Ricci scalar are permitted, namely, $\hat{R}\hat{\Box}\Omega$ and $\hat{G}^{\mu\nu}\hat{\nabla}_{\mu} \hat{\nabla}_{\nu}\Omega$. \\

Integrating by parts the second and third terms, while neglecting total derivatives, leads us to a simplified expression that can be manipulated further. To proceed, we use the following identity:
\begin{align}
\gamma\hat{G}^{\mu\nu} \Bigg(\frac{\hat{\nabla}_{\mu} \hat{\nabla}_{\nu} \Omega}{\Omega}\Bigg)&=\frac{\gamma}{\Omega^2}\hat{G}^{\mu\nu} \hat{\nabla}_{\mu} \Omega \hat{\nabla}_{\nu} \Omega \notag\\
&-\frac{1}{\Omega} \hat{G}^{\mu\nu} \hat{\nabla}_{\mu} \gamma \hat{\nabla}_{\nu} \Omega,
\end{align}
yielding,
\begin{align}
-8\int{\sqrt{-\hat{g}} d^{4}x}\Bigg[\gamma\Omega^{-2}\hat{G}^{\mu\nu} &\hat{\nabla}_{\mu} \Omega \hat{\nabla}_{\nu} \Omega \notag\\
&-\Omega^{-1} \hat{G}^{\mu\nu} \hat{\nabla}_{\mu} \gamma \hat{\nabla}_{\nu} \Omega  \Bigg].
\end{align}

To rewrite the third and fourth terms, we begin by introducing a tensor field defined as follows,
\begin{eqnarray}
J^{\nu}=\hat{\nabla}^{\nu}\Omega \hat{\Box}\Omega -\frac{1}{2} \hat{\nabla}^{\nu}\bigg( \hat{\nabla} \Omega \bigg)^2,
\end{eqnarray}
where $(\hat{\nabla} \Omega)^2\equiv \hat{\nabla}_{\nu} \Omega \hat{\nabla}^{\nu} \Omega$.    With this approach, we ensure that the contributions from the total derivative do not affect the equations of motion, as they will vanish under suitable boundary conditions. Indeed, by focusing on the following total derivative, we can isolate specific contributions from the tensor field  $J^{\nu}$ and facilitate the inclusion of derivatives for the $\gamma$ function as follows,
\begin{align}\nonumber
\hat{\nabla}_{\mu}\Big[\gamma(\Phi)& \Omega^{-2} J^{\mu} \Big]=(\gamma(\Phi) \Omega^{-2}) \hat{\nabla}_{\mu}  J^{\mu}  \\[1ex] 
&-\gamma \Omega^{-2}\bigg[\gamma^{-1} \hat{\nabla}_{\mu} \gamma-2\Omega^{-1}\hat{\nabla}_{\mu} \Omega \bigg]J^{\mu}.\label{boch1}
\end{align}
Eq. (\ref{boch1}) indicates that a convenient scenario can be achieved by choosing $\gamma=\gamma_{0}\Omega^{2}$, provided that the second term cancels out. From this point onward, we shall adopt the latter option primarily for the sake of simplicity. However, we will later demonstrate that this choice holds significant physical relevance. Besides, the divergence term in Eq. (\ref{boch1}) is equivalent to the Bochner identity in curved spacetime,
\begin{eqnarray}
\hat{\nabla}_{\mu}  J^{\mu}=(\hat{\Box}\Omega)^2- \hat{R}^{\mu\nu} \hat{\nabla}_{\mu} \Omega\hat{\nabla}_{\nu} \Omega -  \hat{\nabla}_{\mu} \hat{\nabla}_{\nu} \Omega \hat{\nabla}^{\mu} \hat{\nabla}^{\nu} \Omega.  ~~~~~~~\label{bochx1}
\end{eqnarray}
So far, the contributions from the transformed Gauss-Bonnet (GB) term are given by:
\begin{align}\nonumber
S_{\textbf{GB}}=&\int{\sqrt{-\hat{g}} \,d^{4}x}\bigg[  \gamma \hat{\mathcal{G}} + 8\Omega^{-1}\hat{G}^{\mu\nu} \hat{\nabla}_{\mu} \gamma \hat{\nabla}_{\nu} \Omega   \bigg]+ \\[1ex] 
&\int{\sqrt{-\hat{g}}\, d^{4}x}\Bigg[ -24 \gamma \Omega^{-3} (\hat{\Box}\Omega) (\hat{\nabla} \Omega)^2
+ \notag\\[1ex]
&\qquad\qquad\qquad\qquad 24\gamma \Omega^{-4}   ( \hat{\nabla}\Omega\hat{\nabla} \Omega)^2 \bigg].\label{preq}
\end{align}
We need to express the last integral in Eq. (\ref{preq}) more clearly. To achieve this, the following identities become particularly useful:
\begin{align}
\frac{\hat{\nabla}_{\nu} \Omega \hat{\nabla}_{\mu} \Omega}{\Omega^2} &=\frac{1}{4}\hat{\nabla}_{\nu} \ln \Omega^2 \hat{\nabla}_{\mu} \ln \Omega^2,\\
\frac{\hat{\Box}\Omega}{\Omega} &=\frac{1}{4}\hat{\nabla}_{\nu} \ln \Omega^2 \hat{\nabla}_{\nu} \ln \Omega^2 +\frac{\hat{\Box}}{2}\ln \Omega^2.~~~~\label{preq2}
\end{align}
By replacing Eq. (\ref{preq2}) in Eq. (\ref{preq}), we can combine both terms into a single contribution
\begin{eqnarray}\nonumber
S_{\textbf{GB}}&=&\int{\sqrt{-\hat{g}}\gamma(\Phi) d^{4}x}\Bigg[\hat{\mathcal{G}} + 4\hat{G}^{\mu\nu} \hat{\nabla}_{\mu}  \ln \Omega^2 \hat{\nabla}_{\nu}  \ln \Omega^2 \\ 
&&-3\hat{\nabla}_{\nu} \ln \Omega^2 \hat{\nabla}^{\nu} \ln \Omega^2  \hat{\Box}  \ln \Omega^2 \Bigg].~~~~~~\label{zq}
\end{eqnarray}
At this point, we are led to suspect that there exists a natural choice for parametrizing $\Phi=\Phi(\psi)$. This could be achieved by expressing $\psi$ in the form $\psi=\Omega[\Phi(\psi)]$. After reviewing Eq. (\ref{zq}), we decide to adopt a logarithmic form for this relationship
\begin{eqnarray}
\kappa \alpha\psi= \ln \Omega^{2}[\Phi(\psi)], \label{parame} 
\end{eqnarray}
where $\alpha \in \mathbb{R}$. It is important to note that the parametrization outlined in Eq. (\ref{parame}) is well-established in the literature. For instance, it has been utilized in the context of Higgs inflation, demonstrating its relevance and effectiveness in that framework \cite {Koh:2023zgn,Guo:2009uk,Oikonomou:2020sij}. By choosing  $Q=Q_{0}\Phi^2$ and substituting Eq. (\ref{parame}) into Eq. (\ref{relax}), we can integrate the equation directly. This leads us to the result $\Phi=\Phi_{0}e^{a\psi}$, where  $a$ and $\Phi_{0}$  are constants.   It is noteworthy that any alternative form of $Q$ will give rise to a different relationship between $\Phi$ and $\psi$. In this context, we refrain from assuming a specific form for $Q$ to maintain the generality of our formulation, relying solely on the validity of the relation expressed in Eq. (\ref{parame}).\\

By substituting Eq. (\ref{parame}) into Eq. (\ref{zq}), we derive the equivalent scalarized model in the Einstein frame, where the Ricci and standard kinetic terms are represented in Eq. (\ref{e+k}). However,  the Gauss-Bonnet contribution then takes the following form,
\begin{align}\nonumber
S_{\textbf{GB}}=\int{\sqrt{-\hat{g}}\gamma(\psi) d^{4}x}\Big(\hat{\mathcal{G}} &+ b_{1}\hat{G}^{\mu\nu} \hat{\nabla}_{\mu}\psi \hat{\nabla}_{\nu} \psi\\
&+b_{2}\hat{\nabla}_{\nu} \psi \hat{\nabla}^{\nu}\psi  \hat{\Box} \psi\Big),\label{zq2}
\end{align}
where $b_{1}=4(\kappa\alpha)^2>0$, $b_{2}=-3(\kappa\alpha)^3<0$,  and the global factor is given by
\begin{equation}
    \gamma(\psi)=\gamma_{0}e^{\kappa \alpha \psi}.
\end{equation}
The equations of motion for both the metric and the scalar field can be derived from the work presented in \cite{Koh:2023zgn}. Several comments are in order at this point. First, the equivalent theory in the Einstein frame reveals a specific type of coupling that remains concealed in the Jordan frame. In particular, there exists an interaction between the Einstein tensor and the kinetic term of the scalar field, expressed as $\hat{G}^{\mu\nu} \hat{\nabla}_{\mu}\psi \hat{\nabla}_{\nu} \psi$; such coupling is also characteristic of theories akin to Horndesky types \cite{Babichev:2017guv}. Second, the scalar field acquires a non-linear kinetic contribution, given by the term $\hat{\nabla}_{\nu} \psi \hat{\nabla}^{\nu}\psi  \hat{\Box} \psi$.

The action in Eq. (\ref{zq2}) bears resemblance to the 4D regularized EGB model examined in \cite{Toniato:2024gtx}. However, the connection is not entirely precise, as our model does not include the $(\nabla \psi)^{4}$ term. More significantly, the Gauss-Bonnet term in our formulation is multiplied by $e^{\kappa \alpha \psi}$, whereas in the previous case, it is solely multiplied by $\psi$.\footnote{It is important to note that the contribution from the Gauss-Bonnet action in Eq. (\ref{zq2}) is not necessarily invariant under the parity transformation $\mathbb{Z}_{2}$, characterized by $\psi \rightarrow -\psi$.} Furthermore, the kinetic terms in Eq. (\ref{zq2}) are also accompanied by this exponential factor, which is  absent in \cite{Toniato:2024gtx}. Additionally, the parameter $\gamma_{0}$ governs any deviation from the standard scenario of a scalar field that is minimally coupled to the metric in the Einstein frame. By merely considering a different power of $\gamma=\gamma_0 \Omega^{p}$ with $p>0$, we significantly affect the Lagrangian, leading to the emergence of terms such as $\hat{\nabla}_{\nu} \psi \hat{\nabla}_{\mu}\psi \hat{\nabla}^{\nu} \hat{\nabla}^{\mu}\psi$ [see \cite{Koh:2023zgn} for further details]. One might contend that the exploration of this extension of scalarized models is inherently appealing, particularly since scalar fields coupled to curvature emerge naturally in string models that are ultraviolet complete \cite{Gross:1985rr,Gross:1986mw,Lust:1987xm}. Additionally, a resurgence of interest in stringy gravity models that incorporate higher derivatives has captured considerable attention \cite{ Cano:2021rey, Cano:2022wwo,  R:2022hlf, Daniel:2024lev, Ortega:2024prv, Berti:2025hly}. \\

\subsection{Equivalence Principle}
In GR, it is a fundamental pillar that no preferred frame exists for physical laws, leading to the recovery of special relativity (SR) in the absence of gravity. At the core of GR are two essential principles \cite{Capozziello:2011et,Mancini:2025asp}: the  WEP and the SEP. The WEP asserts that any uncharged free-falling test body, regardless of its composition, follows the same spacetime geodesic. However, its applicability is limited to test bodies in free fall, neglecting the influence of gravity on massive objects \cite{Will:1993ns}. This assertion presents a loophole regarding the universality of free fall in the context of varying couplings in the Einstein frame, in the sense that any differential coupling of the external field to different particle species would indeed challenge the fundamental tenet of the WEP \cite{Damour:1990tw,Damour:1994zq,Damour:2002mi,Bean:2007ny,Damour:2012rc,Bahrami:2018ngq}. In the context of scalar-tensor models, this principle ensures that all matter couples to the Jordan-frame metric, thus establishing the equivalence of inertial and gravitational mass \cite{Damour:1990tw}.\footnote{Notice that the Jordan-frame metric is called the physical one, as it facilitates the recovery of SR within locally inertial frames ($g^{J}_{\mu\nu}=\eta_{\mu\nu}$ and ${}^{J}\Gamma^{\kappa}_{\mu\nu}=0$). Consequently, scalar-tensor theories are structured in a way that they seamlessly embody the Einstein Equivalence Principle.} In contrast, the SEP extends the framework of WEP by affirming that the laws of physics remain consistent across all freely falling frames, including those of massive gravitating bodies. It posits that these laws are independent of the objects' composition or gravitational binding energy, thereby providing a more comprehensive understanding of gravitational interactions \cite{Will:1993ns}.

To wrap up our exploration of the Einstein frame, it is essential to examine the implications of coupling between the metric and the new scalar field, particularly in the context of the geodesic equation. First, we initiate our analysis by considering the conformal transformation of a generic energy-momentum tensor for matter fields that differ from $\psi$ \cite{Damour:1990tw,Casas:1991ky,Faraoni:2020ejh}. Starting with the definition of the EMT in the Jordan frame, we can transition to the Einstein frame as follows,
\begin{eqnarray}
T^{\mu\nu}_{J}=-\frac{2}{\sqrt{-g^{J}}} \frac{\delta S_{\rm{matt}}}{\delta g^{J}_{\mu\nu}}=\Omega^{6} \hat{T}^{\mu\nu},~~~\label{temcon}
\end{eqnarray}
by applying the chain rule along with the identity 
\begin{eqnarray}
\frac{\delta g^{J}_{\mu \nu}}{\delta \hat{g}_{\alpha\beta}}=\Omega^{-2} \delta^{\alpha}_{\mu} \delta^{\beta}_{\nu},
\end{eqnarray}
with $\delta_\mu^\alpha$ the Kronecker delta.
It is established that in the Jordan frame, particles travel along the geodesics defined by the metric $g^{J}_{\mu \nu}$, where the energy-momentum tensor is covariantly conserved, thus  $\nabla_{\nu} T^{\mu\nu}_{{\rm{matt}},J}=0$. By using the aforementioned fact together with Eq. (\ref{temcon}), we can derive the conservation law for the energy-momentum tensor in the Einstein frame. However,  in order to calculate the covariant derivatives, it is essential to incorporate the transformation law of the Christoffel symbols between the two frames \cite{Carneiro:2004rt}, which is given by 
\begin{eqnarray}\label{re1}
{}^{J}\Gamma^{\lambda}_{\mu \nu}=\hat{\Gamma}^{\lambda}_{\mu \nu} + W^{\lambda}_{\mu\nu},
\end{eqnarray}
where the $W$ term can be recast as follows,
\begin{eqnarray}\label{re2}
W^{\lambda}_{\mu\nu}=\hat{g}_{\mu\nu}\hat{\nabla}^{\lambda}\ln \Omega -\delta^{\lambda}_{(\mu}\hat{\nabla}_{\nu)}\ln \Omega^{2}. 
\end{eqnarray}
By combining equations (\ref{re2}) and (\ref{re1}) within the definition of the covariant derivative in the Jordan frame, specifically $\nabla_{\mu} (\Omega^{6} \hat{T}^{\mu\nu}_{\rm{matt}})=0$, we arrive at the following conservation equation in the Einstein frame:
\begin{eqnarray}\label{re22}
 \hat{\nabla}_{\mu} \hat{T}^{\mu\nu}_{\rm{matt}}=- T_{\rm{matt}} \frac{d \ln \Omega}{d \psi} \hat{\nabla}^{\nu}\psi,
\end{eqnarray}
where we have used the fact that $\Omega=\Omega(\psi)$ as previously stated. The physical implications of Eq. (\ref{re22}) warrant several remarks. The nonminimal coupling in the Einstein frame, represented as $\Omega^{2}(\psi)\hat{g}_{\mu\nu}$, signifies an interaction between the scalar field and matter. This implies that the violation of the strong equivalence principle originates from the interaction between the scalar field and the matter Lagrangian. Such a coupling leads to the covariant non-conservation of the energy-momentum tensor, underscoring the energy exchanges that occur not only between matter and the graviton but also with the scalar field. As a result, this interaction can be regarded as a dissipative effect \cite{Damour:1990tw,Casas:1991ky}. This violation of the SEP is also observed in more general scalar-tensor theories, particularly those arising from a disformal transformation (cf.~\cite{Zumalacarregui:2012us} and references therein). One approach to address this phenomenon is to examine the Nordtvedt effect \cite{Nordtvedt:1968qs, Nordtvedt:1968qr}. For example, employing the Lunar Laser Ranging method to measure the distance between the Earth and the Moon \cite{Hofmann:2018myc} enables us to assess whether bodies with differing gravitational binding energies accelerate toward the Sun at varying rates, thus affecting the lunar orbit.\\

At this  point, one might reflect on the conditions under which the universality of free fall, encapsulated by the Weak Equivalence Principle, could be jeopardized \cite{Damour:1990tw,Damour:1994zq,Damour:2002mi}. To illustrate this scenario, we consider two distinct types of matter fields: one associated with dark matter, denoted as $\Psi_{d}$, and another representing baryonic matter, denoted as $\Psi_{b}$. Additionally, we posit that the scalar field $\psi$, which may potentially act as a dark energy candidate, couples in the Einstein frame in a manner that varies according to the nature of the particles or fields involved. Thus, we can express a more general action for the matter field:

\begin{align}
S_{\textbf{T}}= S_{\textbf{R+K}}+S_{\textbf{GB}}&+ S_{\textbf{b}}[\hat{g}_{\mu\nu} e^{\kappa \alpha_{b}(\psi)}, \Psi_{b}]\notag\\
             &+S_{\textbf{d}}[\hat{g}_{\mu\nu} e^{\kappa \alpha_{d}(\psi)}, \Psi_{d}],\label{nmf}
\end{align}
where we used that each sector has a different coupling, $\hat{g}^{\bullet}_{\mu\nu}= e^{\kappa \alpha_{\bullet}(\psi)}g^{J}_{\mu\nu}$. The coupling functions $\alpha_{\bullet}(\psi)$ with $\bullet=b,d$ play a crucial role in defining the interaction strength between the matter sector and the scalar field $\psi$. In this context, the matter sectors include cold dark matter, associated with the coupling function $\alpha_d(\psi)$, and baryonic matter, linked to the coupling function $\alpha_b(\psi)$. This approach closely aligns with previous findings \cite{Damour:1990tw,Bean:2007ny,Bahrami:2018ngq}. The consideration of nontrivial and diverse coupling functions $\alpha_{\bullet}(\psi)$ is well-founded from a theoretical perspective, as it constitutes a common prediction within string theory and models of higher dimensions \cite{Damour:1994zq}.

To connect with our earlier discussion, we examine the action for point-like particles of two distinct species --- dark matter and baryons --- and adopt an alternative approach known as Eardley’s prescription \cite{Eardley:1975fgi}. This framework reveals the emergence of additional terms in the standard geodesic equation for each species within the Einstein frame. A comprehensive treatment of compact bodies in alternative gravity models can be found in \cite{Taherasghari:2022wfs}. We begin with the action for a point-like particle of varying species, denoted by the index $\bullet$:
\begin{eqnarray}\label{app}
S_{\rm{pp},\bullet}=-\int{m_{\bullet}[\alpha_{\bullet}(x^{\mu})] \sqrt{-\hat{g}_{\mu\nu}v^{\nu}v^{\mu}} ds,}
\end{eqnarray}
where we assumed that the mass term $m_\bullet$, is locally dependent on $x^{\mu}$, given that the coupling for each particle fulfills the condition $\alpha_{\bullet}(x^{\mu})= \alpha_{\bullet}[\psi(x^{\mu})]$ (cf. \cite{Damour:1990tw,Damour:1994zq,Damour:2002mi,Taherasghari:2022wfs}).  We define $s$ as the affine parameter that traces the worldline of each particle, with its velocity expressed as $v^{\mu}=dx^{\mu}/ds$ and $\alpha_{\bullet}[\psi]$ denotes scalar invariants \cite{Taherasghari:2022wfs}.  We vary the action while keeping the endpoints fixed and perform a first-order expansion in $\delta x^{\mu}_{\bullet}$. Meaning that we expand   $m_{\bullet}[\psi]$, the proper time $d\eta_{\bullet}/ds=\sqrt{-\hat{g}_{\mu\nu}v^{\nu}v^{\mu}}$, and the four-velocity around the unperturbed worldline of particle with index $\bullet$. 
After integrating by parts and requiring that the first variation of equation (\ref{app}) vanishes for arbitrary variations $\delta x^{\mu}_{\bullet}$, we arrive at the derivation of the modified geodesic equation,
\begin{eqnarray}\label{geo}
v^{\nu}_{\bullet}\hat{\nabla}_{\nu}\big(m_{\bullet}(\alpha_{\bullet}) v^{\bullet~ \epsilon}\big)=-m'_{\bullet}(\alpha_{\bullet})\hat{\nabla}^{\epsilon} \alpha_{\bullet}.
\end{eqnarray}
Alternatively, the expression in (\ref{geo}) can be reformulated in a more convenient manner:
\begin{eqnarray}\label{geox}
v^{\nu}_{\bullet}\hat{\nabla}_{\nu}v^{\bullet~ \epsilon}=-\Big(v^{\nu}_{\bullet}v^{\bullet~\epsilon} \hat{\nabla}_{\nu} \alpha_{\bullet} +\hat{\nabla}^{\epsilon} \alpha_{\bullet}\Big)\frac{\partial \ln m_{\bullet} }{\partial \alpha_{\bullet}}
\end{eqnarray}
Eq. (\ref{geox}) reveals that the spatial variation of local coupling constants results in a violation of the WEP, specifically through the disparities in free-fall accelerations of bodies within an external gravitational field $\psi$. To illustrate this, we consider the non-relativistic limit characterized by $\dot{x}^{i} \simeq \mathcal{O}(v/c)\ll 1$  and $\dot{x}^{0}=1$, in conjunction with the weak field approximation. In this scenario, the metric component is expressed as $g_{00}=-1+2U(x)$, where $U$ represents the Newtonian potential, and the relevant Christoffel symbol is $\Gamma^{i}_{00}=\hat{\nabla}^{i}U$. Long story short, the  modified Newton equation reads,
\begin{equation}
    \ddot{x}^{i}_{\bullet}\simeq -\hat{\nabla}^{i}U_{\bullet} -\frac{\partial \ln m_{\bullet} }{\partial \alpha_{\bullet}}\hat{\nabla}^{i}\alpha_{\bullet},
\end{equation}
indicating the  modified Newton equation for dark matter and baryons are given by 
\begin{align}\label{mn}
 a^{i}_{b}&=g^{i}_{N}-f_{b}\hat{\nabla}^{i}\alpha_{b},\\
    a^{i}_{d}&=g^{i}_{N} -f_{d}\hat{\nabla}^{i}\alpha_{d} \label{mn2},    
\end{align} 
where $g^{i}_{N}=-\hat{\nabla}^{i}U$, $a^{i}_{\bullet}=\ddot{x}^{i}_{\bullet}$,  and the sensitivies of the dark matter/baryon mass on a variation of the coupling read  $f_{b}=\frac{\partial \ln m_{b} }{\partial \alpha_{b}}$  and  $f_{d}=\frac{\partial \ln m_{d} }{\partial \alpha_{d}}$.   Eq. (\ref{mn}-\ref{mn2}) indicate that  there is an anomalous acceleration which depends on the composition of body $d$ and $b$, 
and how they couple to the matter sector at the microscopic level. Consequently, given a microscopic physical model to compute the sensitivities $f_{\bullet}$, any constraint on the universality of free fall can be translated into a constraint on the running of $\hat{\nabla}^{i}\alpha_{\bullet}$.
Since the composition of dark matter is fundamentally different from that of baryonic matter, we can expect that the coefficients associated with the spatial gradients of the coupling constants in (\ref{mn}-\ref{mn2}) will not exhibit universality \cite{Damour:2012rc,Bean:2007ny,Bahrami:2018ngq}, which leads to a violation of the WEP \cite{Capozziello:2011et,Mancini:2025asp}. 
The violation of the WEP has been previously addressed within the context of light scalar dilaton models \cite{Damour:1990tw,Damour:1994zq,Damour:2002mi,Damour:2012rc}, as well as in relation to dark matter field models \cite{Hees:2019nhn}. In our case, 
the relative (dimensionless) acceleration between the two species only involves the sensitives terms \cite{Uzan:2024ded}: 
\begin{align}\label{mn3}
 \eta_{bd}&=\frac{1}{g_{N}}\Bigg|f_{d}\hat{\nabla}^{i}\alpha_{d}-f_{b}\hat{\nabla}^{i}\alpha_{b}\Bigg|,
\end{align}
where $g_{N}=|g^{i}_{N}|$. From (\ref{mn3}) is clear that a non-zero value of $\eta_{bd}$ translates into a violation of the WEP.  
To establish physical constraints on $\eta_{bd}$, it is essential to conduct a thorough analysis of how the external field $\psi$ couples to dark matter and the various matter sectors, specifically electrons and light quarks, at the microscopic level. The aforesaid approach will translate into different running couplings.
This kind of investigation has been explored in the context of dark matter fields in works like \cite{Hees:2019nhn}. 
However, we will set this task aside for now, as it necessitates a thorough examination of a phenomenological microscopic theory \cite{Dent:2008gu, Uzan:2024ded},  which goes beyond the proof of concept presented here.\footnote{The presence of such contributions may pose challenges and it becomes necessary to impose stringent constraints on the fifth force. Alternatively, one might incorporate a screening mechanism that guarantees the effects of the fifth force, which could be significant on large scales, diminish in the regime where GR has been well validated \cite{Khoury:2003aq}. Recently, the notion of circumventing local tests in scalar-tensor theories featuring screening mechanisms has emerged, alongside the detection of scalar field effects in chameleon models utilizing atomic clocks \cite{Levy:2024vyd}.} \\

While the previous discussion offers a profound insight into the intricate relationship between gravity and matter within the framework of sEGB theories, any attempt to impose constraints on violations of the SEP through the Nordtvedt effect — by relying solely on geodesics to articulate the motion of massive bodies — would likely yield limited precision. This limitation arises from the fact that stars and planets are inherently self-gravitating entities, meaning they do not always adhere to the straight paths dictated by geodesics.

To capture a more accurate picture of their dynamics, it is essential to take into account their internal structures and develop equations that describe the motion of each body's center of mass. This complex endeavor becomes more manageable when we employ the traditional post-Newtonian approximation framework and consider the interactions in a system of well-separated bodies. We will explore these intricate details further in the next section.\\

\section{The Nordtvedt effect and observational constraints}
The Nordtvedt effect stands as the most prominent illustration of the violation of the SEP, revealing how self-gravitating bodies can traverse different paths within a gravitational field, influenced by their gravitational energies \cite{Nordtvedt:1968qs,Nordtvedt:1968qr}. In the framework of post-Newtonian approximations, this phenomenon is typically articulated via the quasi-Newtonian acceleration of a body's center of mass, particularly for an $N$-body system. The quasi-Newtonian terms in the acceleration of a body \( A \) are those inversely proportional to the square of the distance and include post-Newtonian corrections related to the body's internal structure (details on Ref. \cite{Will:1993ns}). These can be expressed as modifications to the inertial mass and both passive and active gravitational masses, namely
\begin{equation}\label{mass-tensors}
    (M_{\iota})^{ij}_{A} (a_{j})_A=-(M_p)^{kl}_A\sum_{B\neq A}\frac{(M_a)^{lm}_B}{r^2_{AB}}\,n^l_{AB}n^m_{AB}n^i_{AB},
\end{equation}
where $r_{AB}$ stands for the relative distance between bodies $A$ and $B$, with the unit vector defined as $\bs{n_{AB}}=\bs{r_{AB}}/r_{AB}$. The subindices $\iota,p$ and $a$, attached to $M^{ij}$, indicate the inertial mass, passive and gravitational masses, respectively.

In the context of sEGB theories, as presented in Eq. \eqref{ac1}, these distinct masses align with those found in any fully conservative theory characterized by the PPN formalism, where only the parameters $\beta$ and $\gamma$ deviate from zero. To elucidate this point, we turn our attention to an extended version of the PPN formalism (EPPN), specifically tailored to accommodate sEGB theories. For a comprehensive understanding, we refer to Ref. \cite{sEGB} for further details. The EPPN metric can then be expressed as follows:
\begin{align}
    g_{00}=&-1 +2U - 2\beta_1 U^2  +2\beta_2\Phi_{\cal G} - 2\xi\Phi_W  \notag\\
    &+(2\gamma +2+\alpha_3+\zeta_1-2\xi)\Phi_1\notag\\
    &+2(2\gamma - 2\beta +1+\zeta_2+\xi)\Phi_2 + 2(1+\zeta_3)\Phi_3 \notag\\
    &+2(3\gamma +3\zeta_4-2\xi)\Phi_4 -(\zeta_1-2\xi){\cal A},\label{metricPPN00}\\[2ex]
    g_{0i}=& -\frac{1}{2}(3+\alpha_1-\alpha_2+4\gamma-2\xi+\zeta_1)V_i  \notag\\
    &\quad -\frac{1}{2}(1+\alpha_2+2\xi-\zeta_1)W_i,\label{eppn}\\[2ex]
    g_{ij}=& \ (1+2\gamma U)\delta_{ij}.\label{metricPPNij}
\end{align}
The post-Newtonian potentials ${\cal A},V^i,W^i,\Phi_W$ and $\Phi_1$ to $\Phi_4$ are standard within the PPN framework (see Ref. \cite{Will:1993ns}). Additionally, the potential $\Phi_{\cal G}$ potential is also of post-Newtonian order and it is particular to EGB formulations. Its definition can be found in Ref. \cite{sEGB}, as well as the presentation of the EPPN parameters (the small Greek letters).

The EPPN formulation introduces a significant modification by transitioning the traditional parameter $\beta$ to $\beta_1$ and adding a new parameter, $\beta_2$. This new parameter is essential for capturing the effects associated with the post-Newtonian potential $\Phi_{\cal G}$, which is particularly pertinent to sEGB models. This updated framework is motivated by the recognition that both $\beta_1$ and $\beta_2$ significantly influence the periastron precession observed in binary systems.
By employing this dual parameterization, we can achieve a more refined understanding of gravitational interactions, as it allows us to differentiate the distinct contributions of the modified gravitational potential on the orbital dynamics of these systems. This distinction not only enriches our comprehension of the gravitational behavior within the context of sEGB theories but also enhances our ability to interpret their observational signatures, particularly in relation to phenomena such as periastron shifts in binary systems \cite{Will:1971zzb, Will:1993ns}, which is given by
\begin{align}
    \dot{\tilde\omega} =& ~ 3\left(\frac{2\pi}{P}\right)^{5/3}\frac{M^{2/3}}{1-e^2}\bigg[\frac{1}{3}(2+2\gamma-\beta_1) \notag\\
    &\qquad +\frac{\mu}{6M}(2\alpha_1-\alpha_2+\alpha_3+2\zeta_2) + \frac{J_2R^2}{2Mp}\bigg] \notag\\
    &~\qquad\qquad\qquad\qquad + 6\beta_2\left(\frac{2\pi}{P}\right)^{3}\frac{4+e^2}{(1-e^2)^3}.
    \label{dot-omega}
\end{align}
In the above equation, $P$ represents the orbital period of the system, $e$ is the orbital eccentricity and $p$ stands for the semilatus rectum. The parameter $J_{2}$ captures the influence of the quadrupole moment produced by the dominant body, with $R$ signifying its radius. The system total mass is $M=M_1+M_2$, while the reduced mass is indicated by $\mu$, i.e.,
\begin{equation}
    \mu  \equiv \frac{M_1M_2}{(M_1+M_2)^2}.
\end{equation}

The acceleration of a body's center of mass, as it engages gravitationally with well-separated bodies while accounting for the influence of the potential $\Phi_{\cal G}$, was previously derived in Ref. \cite{Toniato:2024gtx}. Transitioning this result into the EPPN framework is a direct process, yielding the following expression,
\begin{widetext}
\begin{align}
	a_A^j =& \ a_{\s A}^j{\rm \scriptstyle{[PPN]}} + 8\beta_2\sum_{\s B\neq A} \frac{M_{\s B}(M_{\s A}+M_{\s A})}{r_{\s AB}^5} \,n_{\s AB}^j - 6\beta_2 \sum_{\s B\neq A}\frac{E_{\s g,B}^{ik}}{r_{\s AB}^4}\left(\delta^{ik}n^j_{\s AB}+2\delta^{ij}n^k_{\s AB}-5n^i_{\s AB}n^k_{\s AB}n^j_{\s AB}\right)\notag\\[1ex]
	& \ - 4\beta_2\sum_{\s B\neq A}\sum_{\s C\neq A,B}\frac{M_{\s B}M_{\s C}}{r_{\s AB}^2} \Bigg[\frac{n_{\s AB}^j}{r_{\s AC}^3} - 3\frac{(\bs n_{\s AB}\cdot \bs n_{\s AC})n_{\s AC}^j}{r_{\s AC}^3} - \frac{n_{\s BC}^j}{r_{\s AB}r_{\s BC}^2} + 3\frac{(\bs n_{\s AB}\cdot \bs n_{\s BC})n_{\s AB}^j}{r_{\s AB}r_{\s BC}^2} \Bigg].\label{pn-eom}
\end{align} 
\end{widetext}
The term $a_{\s A}^j{\rm  \scriptstyle{[PPN]}}$ denotes the conventional PPN contributions to the equations of motion for each body \cite{Will:1993ns}. It includes the quasi-Newtonian and the post-Newtonian acceleration terms.  Moreover, we introduced $E^{ik}_{\s g,B}$ as the structural integral of body $B$, 
\begin{equation}
	E^{ik}_{\s g,B}=-\frac{1}{2}\int_{\s B}\rs{}'\rs{}''\dfrac{(\bx'-\bx'')^k(\bx'-\bx'')^i}{|\bs\bx'-\bs\bx''|^3}d^3\bx'd^3\bx'',
\end{equation}
with its trace providing a measure of the gravitational energy, $E_{\s g,B}=\delta_{ij}E^{ij}_{\s g,B}$. The conserved density $\rho^{*}$ is defined as the density that satisfies an effective flat-space continuity equation. Eq. \eqref{pn-eom} shows that the modifications introduced by the Gauss-Bonnet potential affect only those terms that decay at a rate of at least $1/r^4$, where $r$ represents the typical distance between the involved bodies. Consequently, the post-Newtonian definitions of inertial mass, as well as passive and active gravitational masses, remain unchanged.

For the sake of simplicity, we restrict our focus to fully conservative theories, wherein the PPN parameters $\alpha$ and $\zeta$ are all set to zero. This scenario is applicable to a majority of alternative gravity theories, including scalar-tensor theories like sEGB \cite{sEGB}. Furthermore, a practical approximation involves treating all bodies as nearly spherically symmetric, which leads to the relation $E_g^{ik}\approx \delta^{ik}E_g/3$. Ultimately, this allows us to simplify the expression \eqref{mass-tensors} for quasi-Newtonian acceleration to
\begin{equation}
    \bs{a_A} =-\frac{(M_p)_A}{M_A}\sum_{B\neq A}\frac{(M_a)_B}{r_{AB}^2}\,\bs{n_{AB}},\label{acc}
\end{equation}
where $(M_p)_A$ denotes the passive gravitational mass of body $A$, while $(M_a)_B$ represents the active gravitational mass of body $B$. In fully conservative theories, these quantities are equal and can be expressed as functions of the gravitational energy as follows \cite{Will_2018}
\begin{align}
    M_p=M_a=M\left[1+\left(4\beta_1-\gamma-3\right)\frac{E_g}{M}\right].\label{mp}
\end{align}
As anticipated, it becomes evident that in GR-like theories, where $\gamma=\beta_1=1$, there is no differentiation between inertial mass and gravitational mass.\\

The most notable implication of the difference between inertial and gravitational masses is observed in the orbital dynamics of a binary system under the influence of a substantially more massive third body. By analyzing equation \eqref{acc}, one can derive the relative acceleration of the binary system, described as \cite{Will_2018}
\begin{align}
\boldsymbol{a}= & -\frac{M^*}{r^2} \boldsymbol{n}-\delta \frac{M_3}{R^2} \boldsymbol{N}+\alpha^* \frac{M_3 r}{R^3}[3 \boldsymbol{N}(\boldsymbol{N} \cdot \boldsymbol{n})-\boldsymbol{n}] \notag\\
& -\frac{3}{2} \Delta^* \frac{M_3 r^2}{R^4}\left[5 \boldsymbol{N}(\boldsymbol{N} \cdot \boldsymbol{n})^2-2 \boldsymbol{n}(\boldsymbol{N} \cdot \boldsymbol{n})-\boldsymbol{N}\right],\label{acc-3b}
\end{align}
with $r$ denoting the separation of the binary system, $\bs n=\bs r/r$, $R$ representing the distance from the third body to the barycenter of the binary system, $\bs N=\bs R/R$, and $M^*$ being the mass of the binary system as measured by Kepler's laws. Moreover,
\begin{align}
\alpha^* & \equiv \alpha+\frac{1}{2} \delta \Delta, \\
\Delta^* & \equiv \Delta \alpha+\frac{1}{2} \delta(1-2 \mu),
\end{align}
where
\begin{align}
\alpha & \equiv \frac{1}{2}\left[\left(\frac{M_{p}}{M}\right)_A+\left(\frac{M_{p}}{M}\right)_A\right] \\[2ex]
\delta & \equiv\left(\frac{M_{p}}{M}\right)_A-\left(\frac{M_p}{M}\right)_B \\[1ex]
\Delta & \equiv \frac{M_A-M_B}{M_A+M_B}.
\end{align}
The first term in Eq. (\ref{acc-3b}) corresponds to the familiar Newtonian acceleration, where $M^*$ signifies the Kepler-measured mass of the binary system; for our purposes, we will substitute it with $M$. The second term accounts for a relative acceleration that either stretches or compresses the orbit along a line directed toward the third body — this phenomenon is known as the Nordtvedt effect, and is mediated by the $\delta$ parameter [cf. \eqref{mp}],
\begin{equation}
\delta=\eta_{\s N}\left(\frac{E_{\s g,A}}{M_A}-\frac{E_{\s g,B}}{M_B}\right),
\end{equation}
where the  Nordtvedt parameter within the EPPN framework can be recast as 
\begin{equation}\label{etan}
\eta_{\s N}=4\beta_1-\gamma-3.
\end{equation}
Eq. (\ref{etan}) tells us that the Nordtvedt parameter is a composite of two PPN parameters  \cite{Will:1971zzb, Will:1993ns}. Besides, any perturbation in the separation of a binary system, attributable solely to the Nordtvedt effect, can be effectively distinguished from conventional tidal effects. This separation allows for precise measurements that can subsequently be utilized to constrain $\eta_{\s N}$.

The Nordtvedt parameter was constrained using different observational methods/systems [cf. Table (\ref{tab:nordtvedt_constraints})]. Therefore, we mention some of the current constraints on  $\eta_{N}$, which will subsequently aid in further constraining the scalarized EGB models within the context of EPPN. One of the earliest stringent constraints on this parameter was established through the LLR experiment \cite{Merkowitz:2010kka}.\footnote{A projection derived from Earth-Mars range measurements, characterized by an accuracy of $\sigma$ meters, is anticipated to yield the following bound for the SEP parameter, $\eta_{N}<(1-12)\times 10^{4}\sigma$ \cite{Anderson:1995yw}.} By measuring the time it takes for light to travel from the observatory on Earth to the mirror on the Moon and back, along with counting the number of detected photons, researchers were able to constrain $\Delta$ for the system Earth-Moon, and determine  $\eta_{N}=(-0.2\pm 1.1)\times 10^{-4}$ \cite{Hofmann:2018myc}. Another significant solar system experiment was the NASA MESSENGER mission, which focused on the determination of Mercury's ephemeris and resulted in the finding that, $\eta_{N}=(-6.6\pm 7.2)\times 10^{-5}$ \cite{Genova:2018mjp}; in fact, a similar estimates should be achieved by the BepiColombo mission \cite{2006AdSpR..38..578M,DeMarchi:2017cbx}. 
 \begin{table}[h]
 \renewcommand{\arraystretch}{1.25}
    \caption{Constraints on SEP  via  $\eta_{N}$.}
    \begin{tabular}{l|l|l}
        \hline\hline
        \textbf{Experiment/Mission} & \textbf{Bound} & \textbf{Ref.} \\
        \hline
        LLR & $\eta_{N}=(-0.2\pm 1.1)\times 10^{-4}$ &\cite{Hofmann:2018myc} \\
        \hline
        MESSENGER  & $\eta_{N}=(-6.6\pm 7.2)\times 10^{-5}$ &  \cite{Genova:2018mjp} \\
        \hline
       BC-MORE &$\eta_{N} < 4.5\times 10^{-5}$ & \cite{DeMarchi:2017cbx} \\  
        \hline
         SR-LP &$\eta_{N} < 2.0 \times 10^{-4}$ & \cite{Congedo:2016mlx} \\ 
        \hline\hline
     %PSR J0337+1715  & $\Delta<2.610^{-6}$ & \cite{Archibald:2018oxs} \\
     %   \hline
     %   PSR J0337+1715 &$\Delta< 0.5\times10^{-6}$ & \cite {Voisin:2020lqi} \\
     %   \hline
    \end{tabular}
    \label{tab:nordtvedt_constraints}
\end{table}

\begin{figure*}[t]
\includegraphics[width=0.4\linewidth]{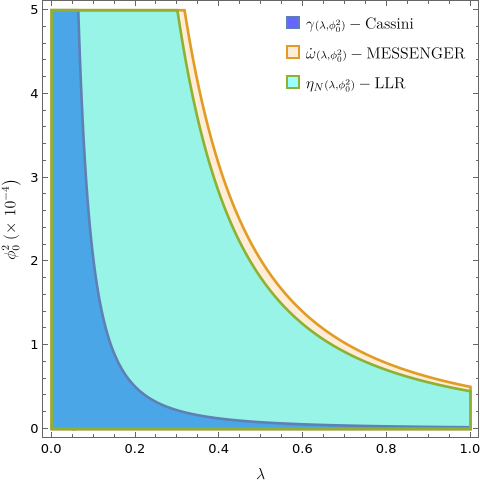}\qquad\qquad
\includegraphics[width=0.4\linewidth]{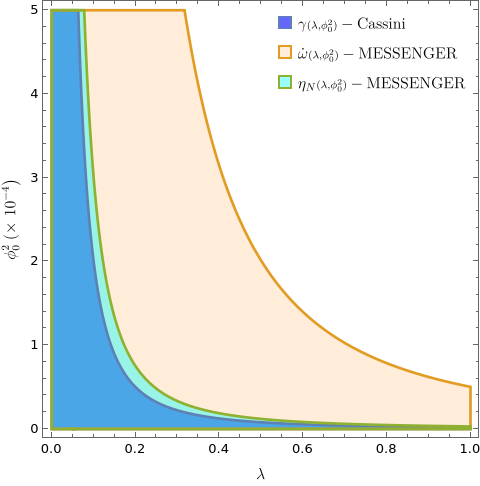}
\caption{In the Ricci-EGB model, we present the combined constraints on the parameter $\gamma$ based on data from the Cassini mission and MESSENGER observations regarding the precession rate of Mercury's perihelion. This is complemented by measurements on the Nordtvedt parameter from LLR (left) and MESSENGER (right).
\label{fig:CML}} 
\end{figure*}

It is well-established that the sEGB model, characterized by $M(\Phi)=Q(\phi)=1$ and $T=0$ , does not induce any modifications to the PPN parameters, given that it is constructed to replicate GR within the solar system context \cite{Doneva:2017bvd}. Therefore, it is essential to consider models in which the PPN parameters deviate from their values in GR. To illustrate this concept, we examine a Ricci-like sEGB model defined by the following coupling functions \cite{Antoniou:2020nax},
\begin{eqnarray}\label{ricci1}
M(\Phi)&=&1-\frac{\lambda}{2}\Phi^{2}, ~~ Q(\Phi)=1,\\
f(\Phi)&=&\frac{\nu}{2}\Phi, \label{ricci2}
\end{eqnarray}
with $\lambda$ and $\nu$ constants. The associated EPPN are given by \cite{sEGB}
\begin{eqnarray}\label{epp1}
\beta_{1}&=&1+\frac{\lambda^2\phi_0^2(2-\lambda\phi_0^2)}{[2+\lambda(4\lambda-1)\phi_0^2]^2[2+\lambda(3\lambda-1)\phi_0^2]},\\[2ex]\label{epp12}
\beta_{2}&=&\frac{-3\lambda\nu \varepsilon\phi_0^2[2+\lambda(2\lambda-1)\phi_0^2]}{64\pi[2+\lambda(4\lambda-1)\phi_0^2][2+\lambda(3\lambda-1)\phi_0^2]},\\[2ex]
\gamma&=&\frac{1}{2}\left[1+\frac{2-\lambda\phi_0^2}{2+\lambda(4\lambda-1)\phi_0^2}\right],\label{epp3}
\end{eqnarray}
and  $\phi_0$ stands for the background value around the post-Newtonian expansion  \cite{sEGB}. To constrain the aforementioned model (\ref{epp1}-\ref{epp3}), we use the currently available bounds of the PPN parameter $\gamma$, Mercury's perihelion shift and the Nordtvedt parameter $\eta_N$.

The Cassini mission measured the frequency shift of radio photons transmitted to and from the spacecraft as it approached the Sun, yielding \cite{Bertotti:2003rm},
\begin{equation}
    \gamma_{\rm{cassini}}-1=(2.1\pm 2.3)\times 10^{-5}.\label{cassini}
\end{equation}
On the other hand, the advance rate of Mercury's perihelion, as reported by the MESSENGER mission \cite{Park:2017zgd}, provides the constraint
\begin{equation}
    \beta_{\rm{perihelion}}-1=(-2.7\pm 3.9)\times 10^{-5}.\label{messanger}
\end{equation}
However, in our exploration of the extended PPN framework, where we redefine the parameter $\beta$ as $ (\beta_1,\,\beta_2)$, it is crucial to consider the implications of Mercury's perihelion shift through the expression given in Eq. \eqref{dot-omega}, whose value observed by MESSENGER is\footnote{In other words,  the constraint \eqref{messanger} can be effectively applied to $\beta_1$ only under the condition that $\beta_2$ is equal to zero, namely  $\beta_1-1=(-2.7\pm 3.9)\times 10^{-5}$.}
\begin{equation}\label{messenger}
    \dot{\tilde{\omega}}_{\rm messenger}=(42.9799\pm 0.0009)\,{\rm arcsec/century}.
\end{equation}
Furthermore, we can utilize constraints on the Nordtvedt parameter (\ref{etan}) derived from the Lunar Laser Ranging  experiment, as detailed in Table \ref{tab:nordtvedt_constraints}. This yields
\begin{equation}
    \eta_{\rm{LLR}}=(-0.2\pm 1.1)\times 10^{-4}.\label{llr}
\end{equation}

Figure \ref{fig:CML} (left) illustrates that the constraints derived from the Shapiro's time-delay phenomenon are significantly more robust than those associated with the precession rate of Mercury's perihelion and the Nordtvedt effect measurements obtained through LLR, particularly at the $1\sigma$ level. Conversely, the MESSENGER mission's bound on $\eta_N$ provides considerably tighter constraints on the parameter space defined by $\lambda$  and $\phi^{2}_{0}$, when compared to Mercury's perihelion shift. However, it still does not match the competitiveness of the constraints obtained from the Cassini mission (see Fig.\ref{fig:CML}). The latter imposes stringent limitations on the parameters $\lambda$ and $\phi_{0}$ for Ricci-EGB models.

It is important to highlight the value of utilizing the Nordtvedt effect to impose additional constraints on the violation of the  SEP within the context of the solar system. Similar results are obtained by taking into account different bounds on $\eta_{N}$ (cf. Table \ref{tab:nordtvedt_constraints}). Looking ahead, we anticipate a deeper exploration of the Nordtvedt effect at the 2PN level in the near future. Indeed, it would be intriguing to investigate, within a certain degree of approximation, the 2PN Nordtvedt effect in the context of a hierarchical triple system. Such an exploration could yield valuable insights into the dynamics of gravitational interactions and the implications for the SEP within scalarized EGB models.

As a closing remark, the violation of the SEP in triple stellar system has been investigated in \cite{Shao:2016ubu,Archibald:2018oxs}, and, particularly within the framework of Bergmann-Wagoner scalar-tensor gravity, in \cite{Voisin:2020lqi}. Utilizing an advanced pulsar timing model alongside six years of data gathered from the Nancay Radio Telescope, researchers demonstrated that a hierarchical triple system constrains the parameter $\Delta$ to be less than $0.5\times10^{-6}$ at the $2\sigma$ confidence level \cite{Voisin:2020lqi}. This bound imposes a substantial condition on the Brans-Dicke parameter, resulting in the significant conclusion that $\omega_{BD}> 140.000$ \cite{Voisin:2020lqi}.

\section{summary}
In this study, we explored the violation of the equivalence principle within a broad class of scalar-Einstein-Gauss-Bonnet theories. We showed that the corresponding model in the Einstein frame introduces a mixed term linking the Einstein tensor to the kinetic terms, alongside higher derivatives in the scalar field, reminiscent of Horndeski-like theories. 
We showed that the resemblance of the equivalent action (\ref{zq2}) to the 4D regularized EGB model \cite{Toniato:2024gtx} is not entirely accurate, as our model omits the $(\nabla \psi)^{4}$ term. Additionally, in our formulation, the Gauss-Bonnet term is scaled by $e^{\kappa \alpha \psi}$  instead of the standard linear scalar field term. Furthermore, the non-linear kinetic terms in our model are also coupled by this exponential factor, which is not present in \cite{Toniato:2024gtx}.

It was shown that in the Einstein frame, the matter action incorporates an exponential coupling mediated by a light scalar field that interacts with all additional matter fields and fluids. To showcase a possible violation of the WEP, we considered a point-like particle framework for dark matter and baryonic matter while employing Eardley's methodology, finding that these particles traverse different geodesics in the Einstein frame when their couplings are not universal or identical.

We proceeded to investigate the violation of the strong equivalence principle by analyzing the Nordtvedt effect using an extended PPN framework. Within this EPPN, the ordinary $\beta$ parameter is represented as $\beta_1$ while $\beta_2$, a new parameter, captures the influence of a new post-Newtonian potential particular to sEGB models. Through this approach, we reviewed how the periastron advance in a binary system is dependent on both $\beta_1$ and $\beta_2$. 

The different components of the acceleration for a binary system affected by a third massive body were identified, which are crucial for a thorough analysis of the violation of SEP and the construction of the Nordtvedt parameter. This latter was then verified to rely on $\beta_1$ and $\gamma$, as in the ordinary PPN formalism.

After presenting a comprehensive summary of the current constraints on the Nordtvedt parameter, as well as predictions from forthcoming space endeavors, we examined a particular sEGB model through a joint constraint analysis leveraging observational data also from the Cassini and MESSENGER missions. While the Cassini constraints on the $\gamma$ parameter (through measurements of Shapiro's time-delay phenomenon) remain the most stringent, it became evident that the limits on the Nordtvedt parameter offer significantly tighter restrictions to the sEGB parameter space than those derived from the precession of Mercury’s perihelion (cf. Fig. \ref{fig:CML}).

In the near future, we aim to enhance our analysis by integrating the Nordtvedt effect at the 2PN level \cite{Damour:1995kt}. While this will necessitate considerable computational resources, it could lead to significant additional constraints on triple systems \cite{Shao:2016ubu,Archibald:2018oxs,Voisin:2020lqi}. It should be necessary to expand our analysis in order to describe the motion of compact objects. Furthermore, exploring chameleon models \cite{Khoury:2003aq} and utilizing atomic clocks \cite{Levy:2024vyd} presents another promising approach to further constrain the fifth force within the framework of an equivalent scalarized EGB model in the Einstein frame.\\

\section{Acknowledgments}
M.G.R was partially supported by WFSA and J.D.T. is grateful for the partial support of FAPES under grant No. 1020/2022.

\bibliography{AllMyRefs}{} 

\end{document}